\documentclass[twocolumn,prb,aps,showpacs,floatfix,superscriptaddress]{revtex4}
\usepackage{graphicx}
\newcommand{\urusi}{URu$_2$Si$_2$}
\newcommand{\dmdt}{\partial M/\partial T}

\begin{document}


\title{Pressure dependence of the magnetization of \urusi}

\author{C. Pfleiderer}
\affiliation{
Physik Department E21, Technische Universit\"at M\"unchen, 85748 Garching, Germany
}

\author{J. A. Mydosh}
\affiliation{
II. Physikalisches Institut, Universit\"at zu K\"oln, Z\"ulpicher Str. 77,
50937 K\"oln, Germany
}

\author{M. Vojta}
\affiliation{
\mbox{
Institut f\"ur Theorie der Kondensierten Materie, 
Universit\"at Karlsruhe, 76128 Karlsruhe, Germany
}}

\date{\today}

\begin{abstract}
The ground state of {\urusi} changes from so-called hidden order (HO) to large-moment
antiferromagnetism (LMAF) upon applying hydrostatic pressure in excess of $\sim$14\,kbar. We report the dc-magnetization $M(B,T,p)$ of {\urusi} for magnetic fields $B$ up to 12\,T, temperatures $T$ in the range 2 to 100\,K, and pressure $p$ up to 17\,kbar.
Remarkably, characteristic scales such as the coherence temperature $T^*$, the transition temperature $T_0$, and the anisotropy in the magnetization depend only weakly on the applied pressure. However, the discontinuity in $\dmdt$ at $T_0$, which measures the magnetocaloric effect, decreases nearly 50\% upon applying 17\,kbar for $M$ and $B$ parallel to the tetragonal $c$-axis, while it increases 15-fold for the $a$-axis. Our findings suggest that the HO and LMAF phases have an astonishing degree of similarity in their physical properties, but a key difference is the magnetocaloric effect near $T_0$ in the basal plane.
\end{abstract}


\maketitle


\section{Introduction}

About twenty years ago two different phase transitions were discovered in \urusi:\cite{pals85,schl86,mapl86} a first transition at $T_0\approx 17.5$\,K, and a second transition at $T_s\approx1.4$\,K to unconventional superconductivity. The transition at $T_0$ leads to a reduction in entropy of about $\Delta S\approx0.2 \rm R \ln 2$. Despite intense experimental and theoretical efforts the ordering phenomenon accounting for this entropy reduction is not understood, hence the phase below $T_0$ in {\urusi} has become known as ``hidden order'' (HO). The mystery of the HO in {\urusi} may be traced to the general complexities of uranium compounds with strong electronic correlations (see e.g. Ref.\,\onlinecite{sces04}). These are (i) the presence of competing energy scales such as strong magnetic anisotropies, strongly hybridized crystal field excitations, and soft lattice modes, (ii) the unknown degree of the itineracy of the three 5$f$-electrons per uranium atom, and (iii) great metallurgical sensitivity.

A large number of microscopic scenarios have been proposed to explain the HO. These
include various versions of spin- and charge-density wave order \cite{maki02,mine05},
forms of crystal electric field polar order \cite{sant94,ohka99,kiss05}, unconventional
density waves \cite{iked98} and orbital antiferromagnetism \cite{chan02}, Pomeranchuk
instabilities \cite{varm06} or nematic electronic phases \cite{barz93}, combinations of
local with itinerant magnetism \cite{okun98} and dynamical forms of order
\cite{fak99,bern03}. None of the models was able to satisfactorily explain all of the
available experimental data; some models are purely phenomenological and lack
material-specific predictions that can be readily verified by experiment, while others
focus only on selected microscopic features.

Let us summarize the key experimental facts. \urusi\ crystallizes in the body centered tetragonal ThCr$_2$Si$_2$ structure. It has long been noticed that the HO exhibits many characteristics of an electronic condensation process: The specific heat anomaly is consistent with a BCS-like gap \cite{mapl86}. The temperature dependence of the resistivity at $T_0$ is strongly reminiscent of the archetypal density-wave system chromium \cite{fawc88}, where slight doping suppresses the anomaly rapidly \cite{kim04}. A change of slope in the finite-field magnetization at $T_0$ suggests the formation of a spin gap \cite{park97}, while optical conductivity indicates a charge gap \cite{bonn88}. Recent thermal conductivity measurements also point towards a gap formation \cite{shar05}.

Early neutron diffraction in the HO phase revealed tiny antiferromagnetic moments of
order $(0.03\pm0.01)\mu_{\rm B}$ per U atom with a $[001]$ modulation and spins aligned along the $c$-axis \cite{broh87}. The magnetic order is clearly three-dimensional with strong Ising-type spin anisotropy. The tiny-moment antiferromagnetism does not, however, account for $\Delta S$ within a local-moment scenario. On the other hand, antiferromagnetism with a rather large moment of $0.4\,\rm \mu_{\rm B}$ and the same Ising-like spin anisotropy pointing along the $c$-axis has been detected in \urusi\ upon applying large hydrostatic pressure $p>p_c\approx 14$\,kbar \cite{amit99}. Numerous experiments indicate as key to an understanding of the HO its relationship with this large-moment antiferromagnetism (LMAF). (Empirically, the average moment of $0.4\,\rm \mu_{\rm B}$ per U atom of the LMAF present at $p>p_c$ would account for $\Delta S$ at $p=0$ in a local-moment picture. Note that no data are available for $\Delta S$ at pressures $p>p_c$.) While neutron scattering cannot distinguish between small homogenous moment and a small inhomogenous volume fraction of large moments, because the scattering intensity is proportional to the sample volume times moment squared, this is not so for NMR and $\mu$-SR, because the signal intensity is directly proportional to the moment. A remarkable piece of evidence are in turn recent neutron scattering \cite{amit99}, NMR \cite{mats01,mats03} and $\mu$SR \cite{amit03b} data which suggest that the tiny-moment antiferromagnetism at ambient pressure represents a tiny volume fraction of LMAF. Based on the these data it therefore appears plausible to assume that the HO is entirely non-magnetic (see e.\,g. Ref.\,\onlinecite{chan03}). However, this issue is controversial, and proposals of HO with an antiferromagnetic component have been made. For instance, it has been pointed out that a spin-density-wave transition close to perfect nesting may exhibit the combination of a small antiferromagnetic moment with a large reduction of entropy \cite{mine05}.

The HO is bounded by more conventional behaviour at high excitation energies as well as at high pressure and high magnetic fields. This is reviewed in the following. Inelastic
neutron scattering in the HO phase \cite{broh91} shows a gap $\Delta(T\!\to\!0)\approx1.8$\,meV in the excitation spectrum on top of the anisotropy gap. At low energies and temperatures, dispersive crystal-field singlet--singlet excitations at the
antiferromagnetic ordering wavevector are observed. These propagating excitations merge above 35\,meV or for $T>T_0$, respectively, into a continuum of quasi-elastic
antiferromagnetic spin fluctuations, as normally observed in heavy-fermion systems. The
excitations exhibit the Ising anisotropy up to the highest energies investigated
experimentally. A rough integration of the fluctuation spectra suggests that the size of
the fluctuating moments would be consistent with $\Delta S$, provided that these moments would be involved in the ordering process \cite{broh91} -- however, it is difficult to draw firm conclusions here.

The application of large magnetic fields parallel to the $c$-axis reduces the
antiferromagnetic signal seen in neutron scattering \cite{maso95,sant00,bour03,bour05}.
At the same time, the ordering temperature $T_0$ collapses to zero at $B_m=38$\,T, and a large uniform magnetization is recovered via a cascade of metamagnetic transitions \cite{harr03,kim03b}. Up to $B_m$ the entropy reduction at $T_0$ stays approximately constant \cite{kim03a}, while the gap $\Delta$, as seen in neutron scattering, increases at least up to 17\,T \cite{bour03}. A topical discussion has been given, e.g., in Ref.\,\onlinecite{harr04}.

In contrast to a magnetic field which destroys the antiferromagnetism, the antiferromagnetic signal is stabilized under uniaxial stress along certain crystallographic directions and hydrostatic pressure. As stated above, NMR \cite{mats01}, $\mu$SR \cite{amit03b} and neutron scattering measurements \cite{amit99} suggest that the system is phase-separated, with the AF volume fraction increasing under hydrostatic pressure and reaching 100\% above $p_c\sim 14$\,kbar. An analogous increase of the AF signal is also seen in neutron scattering under uniaxial stress of a few kbar along the [100] and [110] directions \cite{yoko02,yoko05}, but not under uniaxial stress along the $c$-axis [001]. Inelastic neutron scattering under pressure shows that the dispersive crystal-field singlet excitations at low energies vanish at high pressures \cite{amit00}, consistent with them being a property of the HO volume fraction.

Given the experimental data described so far, the electronic structure near the Fermi
level is of key interest to uncover the nature of the HO. However, a major challenge have
thus far have been direct measurements of the Fermi surface. For instance, de Haas--van Alphen (dHvA) studies under hydrostatic pressure \cite{naka03} do not resolve abrupt changes of the dHvA frequencies and cyclotron masses expected of a distinct phase separation. In these studies the most important observation is a considerable increase of cyclotron mass with increasing pressure.

The nature of the superconductivity in {\urusi} is still little explored, but also provides certain hints on the HO. Early work revealed an unchanged tiny antiferromagnetic moment and was taken to suggest a microscopic coexistence of antiferromagnetism with
superconductivity \cite{broh87,maso90}. The superconducting upper critical field displays an angular dependence under changes of magnetic field direction that can still be explained by Pauli paramagnetic limiting \cite{bris95}. Interestingly, the superconducting transition temperature disappears slowly under pressure well before $p_c $ \cite{mcel87,bris94}, where a reduction of the superconducting volume fraction is also inferred from the magnetization \cite{teny05}. This suggests that the SC is supported by the HO only and cannot coexist with the LMAF.

Finally, many of the controversies around \urusi\ are also related to its complex
metallurgical properties. It has been found that some of the bulk properties are sensitive to the heat treatment the samples received \cite{fak96}. An important impurity effect is the presence of a ferromagnetic component below roughly 30\,K. This may be attributed to a metallurgical impurity phase. Another source of a ferromagnetic impurity signal are stacking defaults of the strongly ferromagnetic planes in the Ising antiferromagnet. For instance, the samples studied recently by thermal expansion under pressure \cite{moto03} contained such a ferromagnetic impurity signal as stated in Ref.\,\onlinecite{uemu05}, but the origin of this signal has not been clarified.

On the theoretical side, the coupling between HO and LMAF has been discussed intensively, given the indications that HO and LMAF constitute the two primary ordering phenomena in \urusi. Phenomenological considerations concern the possible existence of two order parameters (OP), $\psi_{1,2}$, and their mutual coupling in a Ginzburg-Landau framework \cite{shah00}, assuming that both OP are homogeneous over the entire sample volume. Three different possibilities arise: (i) The two OP may break the same symmetries (e.g. lattice translation, time reversal); this would imply that the HO
supports an antiferromagnetic component. Then a linear coupling, $\lambda\psi_1\psi_2$, between the two OP is allowed, and both will be non-zero below an ordering temperature $T_0$. Depending on a microscopic attraction or repulsion of the two OP, the phase diagram will feature a line of first-order transitions between two phases with dominant HO and dominant LMAF order, possibly with a critical endpoint, or a crossover regime only \cite{mine05,bour05}. (ii) The two OP may break different symmetries. Then only a density--density coupling of the form $\lambda'|\psi_1|^2|\psi_2|^2$ is allowed between the two, and there will be phases with one of the OP being zero. Either a first-order or two second-order transitions are required between HO and LMAF. (iii) In certain cases, e.g. $\psi_1$ being a collinear spin-density wave with wavevector $q$ and $\psi_2$ being a charge-density wave with wavevector $2q$, a coupling $\lambda''\psi_1^2\psi_2$ is allowed. Here, $\psi_2$ will always be non-zero once $\psi_1$ is ordered. Experimentally, some evidence has been put forward for a sharp transition at $p_c$ \cite{moto03}, but its nature is under debate. Thermal expansion measurements under pressures are inconclusive as to the existence of a critical endpoint \cite{moto03,uemu05}. We note that neither of these three scenarios has so far been extended to include the possibility of phase segregations over an extended range of pressures.

Clearly, further investigations are required to unravel the relationship between the HO and LMAF phases, in particular comparing the thermodynamic signatures of their transitions -- to our knowledge these have not been comprehensively studied before. Previous experiments under pressure have shown that the onset of either HO or LMAF cannot be distinguished in the electrical resistivity \cite{mcel87,bris95,seki00}. Likewise,
earlier pressure-dependent magnetization measurements hinted at similarities between the two phases \cite{nish00}, but the authors were unable to discern the details of the
transition and overall temperature dependencies. Finally, recent thermal expansion
measurements under high pressure \cite{moto03,uemu05} focussed on the thermodynamic classification of the $T_0$-transition, but failed to provide information of the dominant energy scales or anisotropies.

The purpose of this paper is to close this gap: we report a precision study of the
dc-magnetization of \urusi, where we use hydrostatic pressure as a (nominally) clean
tuning technique to transform the HO into the LMAF. We find that the important energy
scales, notably the transition temperature $T_0$ and $T^*$, being the temperature where $M(T)$ is maximum at fixed external field $B$, remain qualitatively unchanged. The same applies to the magnetic anisotropy. Further, the shape of $M(T)$ around the transition at $T_0$ is also essentially unchanged upon variation of pressure, while the change of slope, i.e., the discontinuity in $\dmdt$ at $T_0$, for $B$ along the $a$-axis is a factor of 15 larger in the LMAF compared to the HO. We conclude that HO and LMAF develop out of very similar disordered states above $T_0$ and have almost identical thermodynamic transition signatures, with the main difference being the in-plane magnetocaloric effect (as measured by $\dmdt$) near $T_0$.


\section{Experimental techniques}

\label{sec:tech}

The single crystal ingot of {\urusi} was grown by means of an optical traveling floating
zone technique at the Amsterdam/Leiden Center. Samples were not annealed, but the optical floating zone method yields comparatively slow temperature reductions that amount to in-situ annealing of the samples. The ingot was characterized by X-ray diffraction and electron probe microanalysis. Bar-shaped samples for the magnetization measurements were sparc eroded from the ingot (typical sample weight 0.05 to 0.1g). The longest part of these bars was aligned parallel to the $c$- and $a$-axis to fit into the pressure cells,
respectively.

We infer the high quality of our samples from their high residual resistivity ratios
(rrr$\approx$20 for the $c$-axis and rrr$\approx$10 for the $a$-axis), the high value of
the superconducting transition temperature ($T_s\approx1.5$\,K) and detailed microprobe analysis which confirmed an excellent stoichiometry and the absence of second phases. Most importantly, however, our samples do not show ferromagnetic inclusions and even for the $a$-axis we see for the first time the transition at $T_0$ and a very shallow Curie tail (see Fig.\,\ref{fig2} below). This contrasts the behaviour of low quality samples \cite{fak96}. Various portions of the same single crystal were used in a variety of other experiments \cite{kim03b}.

The magnetization was measured in an Oxford Instruments vibrating sample magnetometer (VSM) at the University of Karlsruhe. The magnetization of the samples was at first measured at ambient pressure by means of a conventional sample holder. Signal contributions of the sample holder were determined separately and subtracted. The magnetization at high pressures was measured with a bespoke non-magnetic Cu:Be miniature clamp cell, using the same method as reported for previous studies of UGe$_2$ \cite{pfle02}, ZrZn$_2$ \cite{uhla04} Gd$_2$Mo$_2$O$_7$ \cite{kim05}, URhGe \cite{hard05} and CeSi$_{1.81}$ \cite{drot06}. The pressure transmitter was a mixture of
ethanol:methanol (4:1 volume fractions), where we had no indications for stress
anisotropies as suggested in Ref.\,\onlinecite{bour05}. In contrast to studies of
ferromagnetic materials \cite{pfle02,uhla04} the signal strength from the {\urusi} sample
is, however, small. This allowed us to obtain the larger magnetization for the $c$-axis
quantitatively for only a few selected pressures, because careful measurements of the
empty cell were required. Particular features of the magnetization could be tracked as
function of pressure quite easily, notably the maximum at $T^*$ and the kink at $T_0$ at
all pressures. Based on the large number of experiments we have carried out to date, the very slow variation of the signal of the empty pressure cell with temperature, pressure and magnetic field is very well established. Thus, when numerically differentiating the data with respect to temperature, the background from the pressure cell near $T_0$ essentially drops out and a quantitative analysis of the shape of the transition becomes possible. For the magnetization along the $a$-axis we note, that the tiny signal at ambient pressure can not be resolved at all in comparison to typical signal contributions by the pressure cell. However, as discussed below, the signature of the transition at $T_0$ for the a-axis became sufficiently large above 10\,kbar to be visible at high pressure, while the signal was still too small for a reliable quantitative estimate.


\section{Results}

\label{sec:res}

\begin{figure}
\begin{center}
\includegraphics[width=0.38\textwidth]{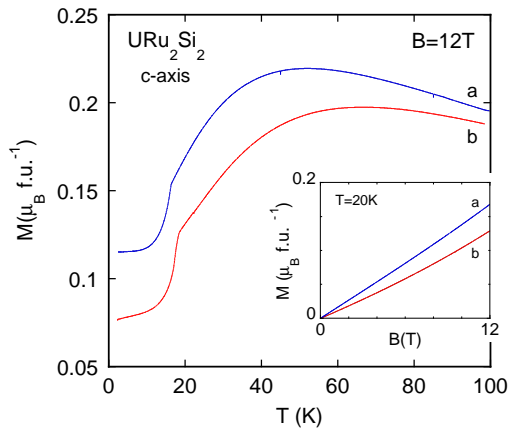}
\end{center}
\caption{
Magnetization $M$ of {\urusi} versus temperature $T$ in the range 2 to 100\,K, for a
field of $B=12$\,T applied along the tetragonal $c$-axis. (a) ambient pressure; (b)
$p=15.9$\,kbar. The inset displays typical data of $M(B)$ for $T=20$\,K, where the
non-linearity in curve (b) is due to the suppression of $T_0$ under magnetic field.
}
\label{fig1}
\end{figure}

At all temperatures and magnetic fields the magnetization of {\urusi} is comparatively
low. In particular, we do not observe any evidence supporting the presence of
meta\-magnetic transitions in the parameter range studied as expected in conventional
local moment antiferromagnets. Typical temperature sweeps for a field along the $c$-axis, both at ambient pressure and at $p=15.9$\,kbar, are shown in Fig.\,\ref{fig1}. For
clarity data are shown for $B=12$\,T. With decreasing temperature a broad maximum at
$T^*\approx50$\,K and 72\,K for $p=0$ and $p=15.9$\,kbar, respectively, is followed by an accentuated drop at $T_0$. The transition temperature $T_0$ decreases under magnetic field along the c-axis from 17.5\,K to around 15\,K. Under pressure $T_0$ and $T^*$ increase weakly. We believe that $T^*$ can be associated with a heavy-fermion coherence scale, as discussed in Sec.~\ref{sec:disc}. The inset displays typical data of $M(B)$ for $T=20$\,K. A linear magnetic field dependence at ambient pressure establishes that the temperature dependence is qualitatively unchanged at all magnetic fields. For the $M(B)$ data at 15.9\,kbar and $T=20$\,K, where $T_0(B=0)\approx$\,21\,K, the suppression of $T_0$ under magnetic field leads to nonlinear contributions in $M(B)$ as shown in the inset of Fig.\,\ref{fig1}.

\begin{figure}
\begin{center}
\includegraphics[width=0.35\textwidth]{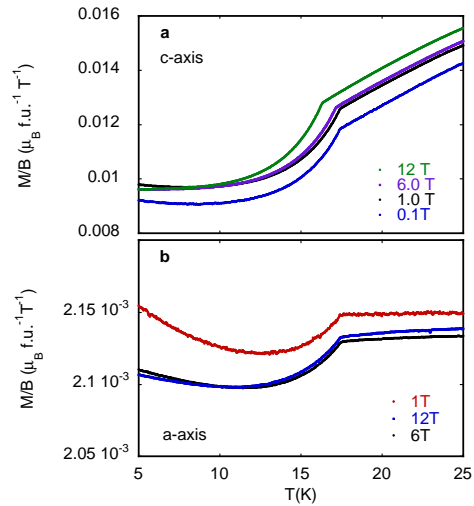}
\end{center}
\caption{
$M/B$ near $T_0$ for the tetragonal $c$-axis, panel (a), and basal-plane $a$-axis, panel (b), at ambient pressure. Data for both directions display a sharp kink at $T_0$ that
signals the formation of a spin gap.
} \label{fig2}
\end{figure}

Figure\,\ref{fig2} shows the remarkably abrupt change of slope in the  $T$ dependence of $M/B$ in the vicinity of $T_0$ at $p=0$ for the $c$- and $a$-axis. $M/B$ is about 5 times smaller for the $a$-axis, consistent with the magnetic anisotropy. The weak upturn for the $a$-axis at 1T signals a very weak ferromagnetic polarisation that we attribute to a tiny number of defects, i.e., it strongly supports a very high sample quality (cf.
Ref.\,\onlinecite{fak96}). We note that the transition at $T_0$ to our knowledge has not
been seen for the $a$-axis in $M(T)$ before. We have carefully confirmed experimentally that the signal measured for the $a$-axis is not contaminated by any possible $c$-axis contributions. The conclusion that we measure indeed purely the $a$-axis finds further support by the difference of magnetic field dependence of $T_0$ for the $a$-axis and $c$-axis (cf. Fig.\,\ref{fig4} below).

\begin{figure}
\begin{center}
\includegraphics[width=0.35\textwidth]{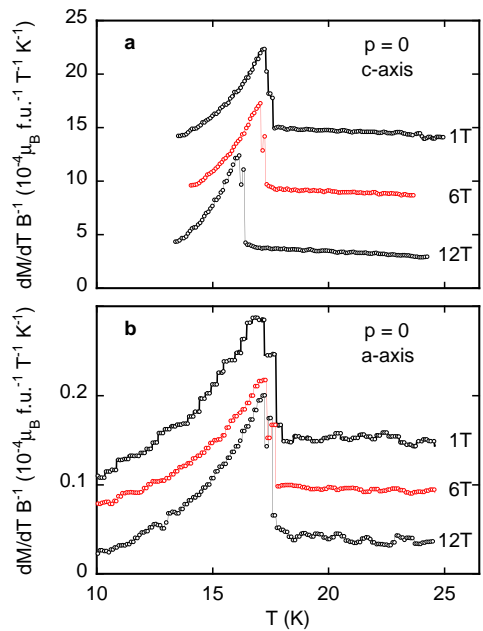}
\end{center}
\caption{
Typical numerical derivative $\dmdt$ vs $T$ of $M(T)$ measured experimentally for various fields at ambient pressure for the $c$-axis, panel (a), and the $a$-axis, panel (b). The discontinuous steps are purely due to the method of calculation. (Note that the data are normalized to the applied field, and the curves have been shifted by arbitrary constants for clarity.)
}
\label{fig3}
\end{figure}

To track the change of slope of $M(T)$ at $T_0$ as function of $p$ we have computed the derivative $\dmdt$ from the experimental data, where the gentle numerical smoothing introduces the raggedness of the data points shown in the figures. In particular, consideration of possible evidence supporting a double in $T_0$ for the basal plane requires extensive additional studies. The discontinuity of the derivative provides also a measure of the magnetocaloric effect, $\dmdt\vert_H=\partial S/\partial H\vert_T$. Derivatives of the ambient pressure data are shown in Fig.\,\ref{fig3}. The shape of the derivative is qualitatively very similar for both field directions. Quantitatively, the discontinuity between the $c$ and $a$-axes are different by nearly a factor of 50.

The derivative $\dmdt$ calculated from the measured magnetization at high pressure is
shown in Fig.\,\ref{fig4}. For both directions the qualitative shape of the $\dmdt$ curve
near the transition is nearly field-independent. Based on our data we observe no evidence suggesting a change of the transition from second to first order or similar.  Quantitatively, we observe that the size of the discontinuity in $\dmdt$ decreases moderately upon applying pressure for fields $B$ along the $c$-axis, while that for $B$ along the $a$-axis increases nearly 15-fold between ambient pressure and 17.2\,kbar. This indicates that a crucial difference between HO and LMAF is the magnetocaloric effect in the basal plane. Further we note that with increasing magnetic field along the $c$-axis the transition develops additional structure. To our knowledge this is the first evidence for additional sub-phases of the LMAF. Interestingly, the total height of the peak in $\dmdt$ remains unchanged even in the presence of the additional structure. It is possible that the origin of the double transition are small pressure anisotropies in our pressure cell. However, (i) all of the previous studies \cite{pfle02,uhla04,kim05,hard05,drot06} suggested excellent pressure homogeneity, and (ii) the anisotropy would be expected along the pressure cell, but uniaxial stress studies show that {\urusi} is insensitive to stress along the c-axis.

An important aspect, already visible in Fig.\,\ref{fig4}, is the field-dependence
of $T_0$, shown in Fig.\,\ref{fig5} for both the $a$ and $c$-axes for $p=0$ as well
as for high pressure. The value of $T_0$ is defined at the onset of the transition upon decreasing temperature. (Note that the additional structure seen at high pressure for the $c$-axis is not reflected in this plot.) A magnetic field along the $c$-axis suppresses $T_0$, in agreement with previous studies \cite{harr03,kim03a,kim03b}. It appears that $T_0(B)$ for the $c$-axis drops slightly faster at high pressure. However, the faster relative drop may be traced to the increase of $T_0$ under pressure, i.e., $T_0(B)$ is quantitatively unchanged despite the increase of $T_0$. Most remarkably, for $B$ along the $a$-axis the transition temperature $T_0$ is {\em not} affected by the field (up to 12\,T), both at ambient pressure and at high pressure. Thus, even though the state below $T_0$ changes from HO at ambient pressure to LMAF at high pressure, the variation of $T_0(B)$ is {\em quantitatively} unchanged. This indicates that both phases share a remarkably similar transition mechanism.

\begin{figure}
\begin{center}
\includegraphics[width=0.35\textwidth]{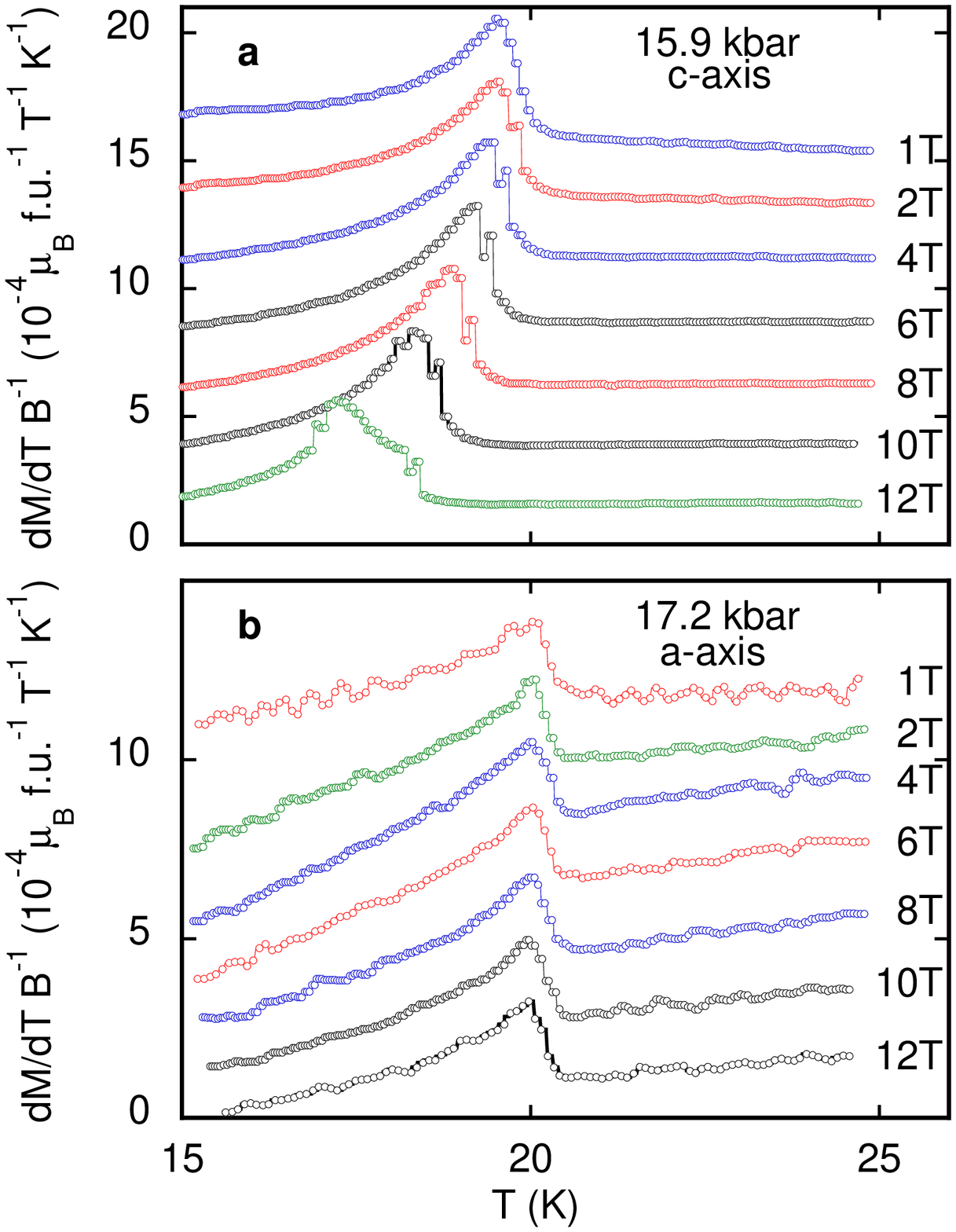}
\end{center}
\caption{
Typical numerical derivative $\dmdt$ vs $T$ of $M(T)$ measured for various fields at high pressure for the $c$-axis, panel (a) and for the $a$-axis, panel (b). For the $c$-axis
additional structure, that emerges with increasing magnetic field, is seen. This indicates at a more complex magnetic state at high pressure. (As in Fig.\,\ref{fig3}, the data are normalized to the applied field, and the curves have been shifted by arbitrary constants for clarity.)
}
\label{fig4}
\end{figure}

Key features of our magnetization data are summarized in the phase diagram,
Fig.\ref{fig6}. Panel (a) displays the pressure dependence of $T^*$, which increases
weakly with pressure from 50 to 75\,K. Likewise, the transition temperature $T_0$
increases only slightly with pressure as shown in Fig.\,\ref{fig6}\,(b), with a change of
slope around $p_c\approx 14\,$kbar. The pressure dependence of $T_0$ we observe is
consistent with previous resistivity and neutron scattering measurements.
Fig.\,\ref{fig6}\,(c) brings out key differences of the thermodynamic signatures of $T_0$
when going from nearly 100\% volume fraction of HO to 100\% volume fraction of LMAF. The change of slope for field along the $c$-axis drops moderately by roughly 30\,\%. In
contrast, the change of slope along the $a$-axis increases 15-fold, where the data point
for $p=0$ was measured without pressure cell. As stated above, it is not possible to
resolve the low value of $\dmdt$ at $p=0$ when the sample is measured together with the pressure cell. This also explains why no transitions could be detected for 4\,kbar and
8\,kbar, respectively, i.e., for these pressures  $\dmdt$ must be still very low. Yet,
the large high-pressure value of $\dmdt$ for the $a$-axis is observed for all $p\geq
12\,$kbar and thus already below $p_c$. We note that a relatively steep transition line
between HO and LMAF has been reported in recent neutron scattering studies \cite{bour05} that contrasts earlier neutron scattering studies under pressure \cite{amit99}.

We have no evidence that the anisotropy in the magnetic susceptibility changes
substantially under pressure. Thus the environment in which the transition at $T_0$ takes place is essentially unchanged, even though neutron scattering, NMR and $\mu$SR show that the microscopic characteristics of the ordered phases radically change under pressure. To our knowledge, the magnetization provides the first thermodynamic evidence in clean samples that the HO and LMAF have essentially identical transition properties.

\begin{figure}
\begin{center}
\includegraphics[width=0.38\textwidth]{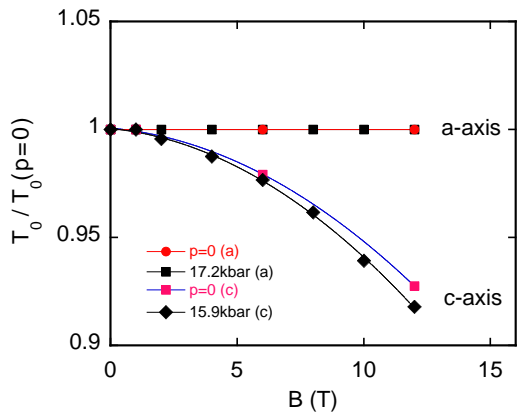}
\end{center}
\caption{
Normalized field dependence of $T_0$ at ambient pressure and high pressure (above $p_c$) for the $c$ and $a$ axes. As transition temperature $T_0$ we take the onset of the transition upon decreasing temperature. The additional structure for the $c$-axis at
15.9\,kbar and 12\,T shown in Fig.\,\ref{fig4}(b) is not represented here. We find that
the suppression of $T_0$ with $B$ along the $c$-axis is unchanged upon applying pressure, as discussed in the text.
}
\label{fig5}
\end{figure}

\begin{figure}
\begin{center}
\includegraphics[width=0.42\textwidth]{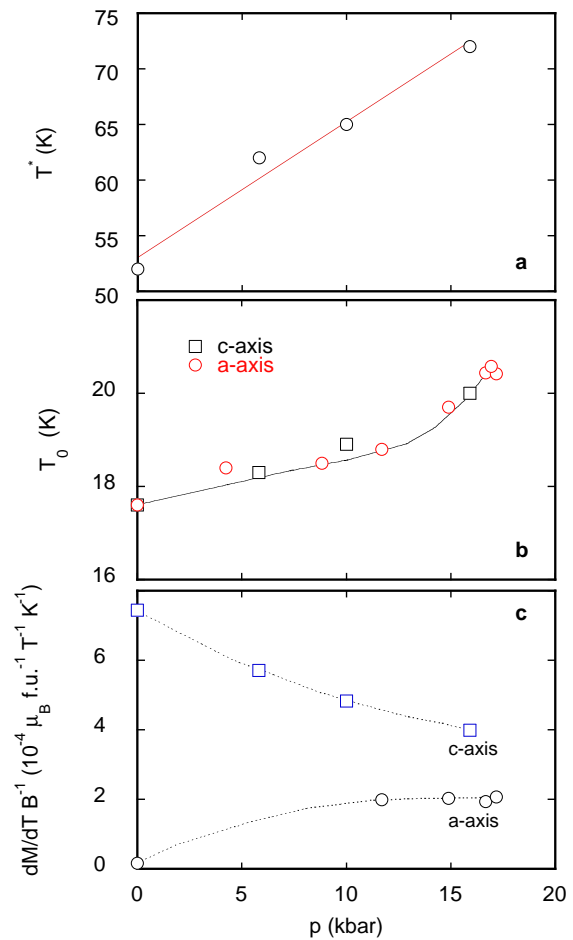}
\end{center}
\caption{
Pressure dependencies of the coherence temperature $T^*$, the transition
temperature $T_0$, and the total height of the anomaly in $\dmdt$. Lines are guides to
the eye. (a) The temperature $T^*$ of the maximum in $M(T)$ increases nearly 30\% under pressure up to $\sim$20\,kbar. (b) The transition temperature $T_0$ increases weakly under pressure, with a pronounced change of slope around 14\,kbar. The change of slope suggests that the border between HO and LMAF is crossed, and may be used to define $p_c\approx 14\,$kbar. (c) The total height of the anomaly at $T_0$ seen in $\dmdt$. For the $c$-axis $\dmdt$ decreases by nearly 50\%, while $\dmdt$ \textit{increases} nearly 15-fold for the basal plane $a$-axis. It is interesting to note that the height of the anomaly $\dmdt$ for the $a$-axis is already maximal \textit{below} the $p_c$ and remains constant between 12 and 17\,kbar. The data point for the $a$-axis at $p=0$ was recorded without pressure cell (cf. Fig.\,\ref{fig3}).
}
\label{fig6}
\end{figure}


\section{Discussion}
\label{sec:disc}

We now discuss our experimental findings. In our study we have established that both the behavior of the bulk magnetization near the transition at $T_0$ and temperature
dependence $T_0(B)$ remain qualitatively unchanged upon increasing the pressure from 0 to 17 kbar. Yet, it is known that the ordered state microscopically changes dramatically under pressure when going from HO to LMAF. A plausible conclusion is that the HO and LMAF share a very similar ordering mechanism, albeit with different order parameters. The manner how the HO changes into LMAF under pressure is an issue of great current controversy (see e.g. Ref.\,\onlinecite{bour05}), which we cannot directly access from our data. We note, however, that we do not see any signs in our data of the location of the cross-over or transition line between HO and LMAF.

A few remarks regarding the magnetism are therefore in order. In the parameter range of our experiments we do not observe any evidence for metamagnetic transitions characteristic for local-moment magnets: At least in the vicinity of $T_0$ our experimental setup should be sensitive to such a metamagnetic transition, provided that the moment is of order 0.4 $\mu_{\rm B}$ for $T\to0$. In comparison to the energy scale set by $T_0$, the variation of $T_0$ under magnetic field is very weak for the $c$-axis and even absent for the basal plane (which appears inconsistent with local-moment magnetism). Notably, the behaviour for the $a$-axis observed here is similar to the absence of magnetic field dependence of spin-density wave order in Cr \cite{fawc88} and Mn$_3$Si \cite{pfle02b}. In combination with the features of the resistivity, specific heat and many other properties, this again strongly suggests that the magnetism is of itinerant character, and may be understood as a condensation process of the conduction electrons at $T_0$.

The maximum of $M(T)$ at a temperature $T^*$ being significantly larger than $T_0$
(Fig.~\ref{fig1}) points towards the presence of heavy-fermion physics: In a Fermi liquid there is a $T^2$ correction to the $T=0$ Pauli susceptibility which may have positive sign; in contrast, in a paramagnetic local-moment system the magnetization should decrease with increasing $T$. Kondo-screened moments, with $T^*$ being a measure of the Kondo (or coherence) temperature, will be quenched below $T^*$ and thus can possibly account for the data. Thus, it appears reasonable to assume that the transition at $T_0$ appears within a (partially formed) heavy-fermion state.

An interesting empirical observation is that the experimental variation of $\dmdt$ near
$T_0$ is strongly reminiscent of the specific heat anomaly. In fact, we find that $\dmdt$
and $C(T)$ closely track each other near $T_0$, i.e., they are proportional to one
another. This suggests that the spin excitations are indeed the degrees of freedom
dominating the spectrum of excitations at $T_0$, with the pronounced drop of $M(T)$ below $T_0$ signaling the formation of a spin gap. Unfortunately it is not possible to relate the magnetocaloric effect $\partial S/\partial B$ in general terms to the specific heat $C=\partial S/\partial T$ without use of a specific model.  In the comparison of the HO and LMAF this leaves as a major challenge for future studies the specific heat anomaly at $T_0$ under pressure.

We also note the similarity of $\partial M/\partial T$ and the non-linear susceptibility
$\chi_3$ reported by Ramirez {\em et al.} \cite{rami92}. Like the specific heat, $\chi_3$
tracks $\dmdt$. For the $c$-axis this similarity may be related to the weak reduction of
$T_0$ under field which leads to a very weak non-linearity of $M(B)$ up to 5\,T, the
field range studied in Ref.\,\onlinecite{rami92}. Yet, as shown in Fig.\ref{fig5}, there
is no suppression of $T_0$ under field for the $a$-axis, but $\dmdt$ at high pressure
becomes comparable in size to the behavior seen for the $c$-axis. When taken together
this questions the uniqueness of the interpretation of $\chi_3$ given in
Ref.~\onlinecite{rami92}.

We wish to return to aspects of the electronic structure of \urusi. It is interesting to consider parallels to the itinerant ferromagnets UGe$_2$ \cite{pfle02} and ZrZn$_2$ \cite{uhla04}: In both compounds changes of the magnetization at itinerant metamagnetic
transitions may be explained in terms of Fermi surface reconstructions, when a Fermi surface sheet is driven across the boundary of the Brillouin zone under hydrostatic pressure and magnetic field. In the case of \urusi, the tetragonal crystal structure suggests cylindrical Fermi surface sheets parallel to the $c$-axis. The high sensitivity to changes of the basal-plane lattice constant, seen under uniaxial pressure \cite{yoko02,yoko05}, provides support of a Fermi surface instability akin UGe$_2$ and ZrZn$_2$. (We note that the rather large value of $p_c$ is not inconsistent with the notion of
HO and LMAF being almost degenerate in free energy, as the $a$-axis lattice constant
under hydrostatic pressure decreases only weakly \cite{devi86,vand95}.) The scenario of cylindrical Fermi surface sheets, together with the $c$ axis being spin easy axis, is finally also consistent with the lack of field dependence of $T_0$ for fields along the $a$-axis, provided that the transition is driven by features in the density of states.

Unfortunately, measurements of the Fermi surface topology in {\urusi} so far have been
inconclusive. Only tiny portions have been observed, and strong damping of the heavy-fermion bands is seen under pressure when entering the LMAF state \cite{naka03}. At the same time, the reported increase of the cyclotron mass under pressure, as inferred from dHvA studies, raises another issue: If the HO and LMAF are strictly phase-segregated, this mass enhancement may no longer be related to the properties of a single band in the conventional approach.

Let us turn to phenomenological theoretical considerations.
Our measurements of $M(T,B)$ reflect the coupling between the order parameter
(of the HO or LMAF phase) and the static uniform magnetization $M$.
One may consider a Landau functional for the free-energy density $f$
of the form
\begin{eqnarray}
f =&&
a(T) |\psi|^2 + b |\psi|^4 +
\sum_\alpha v_\alpha M_\alpha^2 |\psi|^2 \nonumber\\
&&+
\sum_\alpha (g_\alpha M_\alpha B_\alpha + u_\alpha M_\alpha^2)
\end{eqnarray}
where $\alpha=a,b,c$ denotes the three axes,
$\psi$ is the order parameter of the phase below $T_0$,
$a$ and $b$ are the coefficients of the standard Landau expansion of $\psi$,
$v_\alpha$ is the coupling between $M$ and $\psi$,
$g_\alpha$ is the $g$ tensor in diagonal form,
and $u_\alpha$ is the quadratic Landau term for $M$.
It is easy to see that the magnetization below $T_0$ is given by
$M_\alpha=g_\alpha B_\alpha / (2 u_\alpha + 2 v_\alpha |\psi|^2)$,
whereas the change of $T_0$ as function of $B$ is proportional
to $g_\alpha^2 v_\alpha/u_\alpha^2$.
Thus, the data may be consistent with $v_a$ increasing significantly and
$v_c$ decreasing somewhat under pressure,
$g_c^2 v_c$ being roughly pressure-independent,
and $g_a$ being small enough that the change in $T_0$ caused by $g_a^2$
is negligible (assuming $p$-independent $u_\alpha$).
This suggests a strong difference of HO and LMAF regarding their
in-plane magnetic properties.

We finally comment on the theoretical models proposed for \urusi. Purely phenomenological models of two order parameters, treated on the mean-field level, are not able to consistently describe $M(T)$ as found in experiment. Mineev and Zhitomirsky\cite{mine05} have investigated a model for local uranium moments (causing LMAF) coupled to a spin-density wave order parameter with small form factor (causing HO). While this model seems to reproduce some important features of \urusi, we do not expect it to fully account for the behavior of $M(T)$: The crystal-field physics of the model does not easily reproduce a maximum in the magnetization at $T^*$ -- we think that this requires Kondo screening of the moments (which is not included in the model of Ref.~\onlinecite{mine05}) to set in around $T^*$. Also, it is open how the very similar transition properties at ambient and high pressure would emerge from this model. Similar arguments apply to the scenario of ``helicity order'', put forward by Varma and Zhu \cite{varm06}. They propose a $p$-wave spin-triplet Pomeranchuk instability as candidate for the HO. So far they have not included microscopic features like the Fermi surface topology, nor did they consider the interplay of the HO with the LMAF at high pressure. Also, in its present form the model apparently fails to explain the experimentally observed formation of a gap below $T_0$.

Clearly, fresh theoretical input is needed: We believe that the concept of two order parameters combined with both crystal-field and heavy-fermion physics is required to fully describe the properties of the magnetization for the whole pressure range. Furthermore, any candidate scenario should account for the operning of a (perhaps partial) electronic gap at the $T_0$ transition.


\section{Conclusions}
\label{conc}

We have studied the magnetization $M(T,B,p)$ in \urusi\ single crystals. Although $M$
does not directly couple to the dominant order parameters of both the HO and LMAF phases, we can draw a number of important conclusions from our experimental results: Pressures up 17 kbar increase the coherence temperature $T^*$ by almost 50\%. This means both the HO and LMAF states evolve out of a partially formed heavy Fermi liquid where the high-$T$ local moments of the U-ions are almost screened. The drop of the magnetization $M(T)$ below $T_0$, with the sharp, mostly field independent, knees of $M/B$, indicate the opening of a spin gap. This effect is much larger for fields applied along the $c$-axis than for those along $a$. In order to quantify this behavior we have determined $(1/B)\dmdt$ as a function $B$ and $p$ for $T$ near $T_0$, which is related to the excitation spectrum at the HO and LMAF transitions. The qualitative form of $\dmdt$ does not change between these two states, similar to the signature of the resistivity at $T_0$ which also changes little with pressure \cite{mcel87}.

One important difference between HO and LMAF uncovered in our study is the anisotropy in $\dmdt$: pressure causes the peak in $(1/B)\dmdt$ to be reduced for the $c$-axis, yet it is increased for the $a$-axis. High pressure thus removes the large anisotropy in $\dmdt$ between these axes, which is present at ambient pressure.
The pressure-invariant transition signatures around $T_0$ in the observables listed above show that HO and LMAF are not only phases which are almost thermodynamically degenerate (i.e., have almost the same free energy density), but they also have little difference regarding their transitional properties. We conclude that HO and LMAF must evolve out of the same related physical ingredients.

In our view, none of the available theoretical scenarios can easily explain our
experimental findings. On the experimental side, to gain further information of the
pressure dependencies specific heat offers the best possibility. For, by combining
$(\dmdt)(T, B, P)$ with $C_p(T, B, P)$ we can determine the full magnetocaloric effect
and thus the entropy and Gr\"uneisen parameter, $(T^{-1} dH/dT)_S$, of the pressure
transformation from HO to LMAF.


\acknowledgments

We have benefited from stimulating discussions with P. B\"oni, B. F{\aa}k, C. M. Varma,
and in particular with A. Rosch. CP was supported in the framework of a Helmholtz-Hochschul-Nachwuchsgruppe during the initial part of this project. Financial support by the Alexander von Humboldt foundation (JAM) and the Virtual Quantum Phase Transitions Institute Karlsruhe (MV) is also gratefully acknowledged.

\bibliography{references}

\end{document}